# Flute-Model Acoustic Metamaterials with Simultaneously Negative Bulk Modulus and Mass Density

H. C. Zeng, C. R. Luo, H. J. Chen, S. L. Zhai and X. P. Zhao[*]

Smart Materials Laboratory, Department of Applied Physics,
Northwestern Polytechnical University, Xi'an 710129, People's Republic of China
E-mail: xpzhao@nwpu.edu.cn

## Abstract

We experimentally constructed a three-dimensional flute-model "molecular" structure acoustic metamaterial (AM) from a periodic array of perforated hollow steel tubes (PHSTs) and investigated its transmission and reflection behaviors in an impedance tube system. The AM exhibited a transmission peak and an inverse phase, thus exhibiting the local resonance of the PHSTs. Based on the homogeneous media theory, the effective bulk modulus and mass density of the AM were calculated to be simultaneously negative; the refractive index was also negative. PHST AM slab focusing experiments showed that the medium with a resonant structure exhibited a distinct metamaterial property.

---

[*]To whom correspondence should be addressed.   E-mail: xpzhao@nwpu.edu.cn   Tel: 86-29-88431662

In 1968, Veselago[1] theoretically discussed artificial periodic structural left hand metamaterials (LHMs) with simultaneously negative permittivity ε and permeability μ based on Maxwell's equation. In 2001, Smith experimentally fabricated LHMs in the microwave frequency band[2–5] by combining an array of wires with split resonant rings (SRRs) and caused widespread concern over LHMs. An LHM can display a series of unique "reversed" characteristics, including negative refraction, sub-wavelength focusing, sub-wavelength imaging, and invisibility cloaking.[6–12] Novel phenomena have changed our knowledge of traditional materials such that various new applications previously unimagined can now be achieved.

Given the similarity of electromagnetic and acoustic waves, the application of an acoustic metamaterial (AM) has attracted significant research attention. Sheng et al.[13] proposed the concept of local resonance, first realizing a negative mass density in an acoustic field and then achieving a resonant effect on sound waves of different frequency bands by designing the geometry of a resonant microstructure. Fang et al.[14] found a negative effective dynamic modulus near the resonant frequency in a structured composite consisting of a one-dimensional array of repeating unit cells with shunted Helmholtz resonators (HRs). Ding et al.[15–17] established an split hollow sphere(SHS) microstructure to achieve a negative bulk modulus. To further explore double-negative AMs, the current report proposes a perforated hollow steel tube (PHST) flute model with a "molecular" resonant structure. The transmission and reflection behaviors of this model are then investigated in an impedance tube system. The AM exhibited a transmission peak and an inverse phase. Based on the homogeneous media theory, the effective bulk modulus and mass density of the AM were calculated to be simultaneously negative; its refraction index was also negative. Through PHST AM slab focusing experiments, we found that the medium with a resonant structure exhibited a distinct metamaterial property.

LHM analogues, which combine an array of wires with SRRs, were experimentally fabricated at the microwave frequency band.[2,3] Ding et al.[15–17] constructed an SHS model that could achieve a negative bulk modulus. Our research group proposed the HST model, which can achieve a negative effective mass density. Based on these models, we decided to perforate the HST as shown in Figs. 1(a) and 1(b).

A PHST, a type of hollow steel tube that is open at both ends, was perforated 1/3 away from one of the steel tube ends. The opening at both ends of the cylindrical hollow steel tube can guide a sound wave. This structure is equivalent to the inductance of an acoustic circuit, $L_t = \rho_0 l_t / S_t$, where $S_t$ is the cross-sectional area of the steel tube end and $l_t$ is the effective length of the tube port. The tube port is a hollow tube cavity that stores sound wave energy, which is equivalent to the acoustic volume of an acoustic circuit, $C_t = V_t / (\rho_0 c_0^2)$, where $V_t$ is the volume of the steel tube cavity, $\rho_0$ is the density of the fluid, and $c_0$ is the speed of sound in the fluid. The hole in the tube is equivalent to the neck of an HR, which acts as the acoustic inductance, $L_h = \rho_0 l_h / S_h$. At the same time, the hollow cavity is equivalent to the acoustic volume, $C_h = V_h / (\rho_0 c_0^2)$, where $l_h$, $S_h$, and $V_h$ are the thickness of tube wall, the diameter of the hole, and the volume of hollow cavity, respectively. The PHST is equivalent to the acoustic "molecule" consisting of the acoustic "atoms," SHS and HST. Based on L-C resonance model in Fig. 1(c), we can obtain the resonance frequency, $f = \dfrac{1}{2\pi\sqrt{L_{eff}C_{eff}}}$.

The PHST was fabricated with a radius of $r_1$=19 mm, tube wall thickness of $d$=0.3 mm, and perforated hole $r_2$=3 mm positioned 1/3 away from one of the steel tube ends. We first constructed a 40 mm thick sponge matrix with a 100 mm radius. Then, we periodically alternately arrayed the PHSTs in a zigzag manner at both upper and lower surfaces relative to the location of the hole. A double-layered PHST sample A was fabricated, as shown in Figs. 1(d) and 1(e), to enhance the resonance effect. To compare with the AM, a plain sponge sample B was also fabricated.

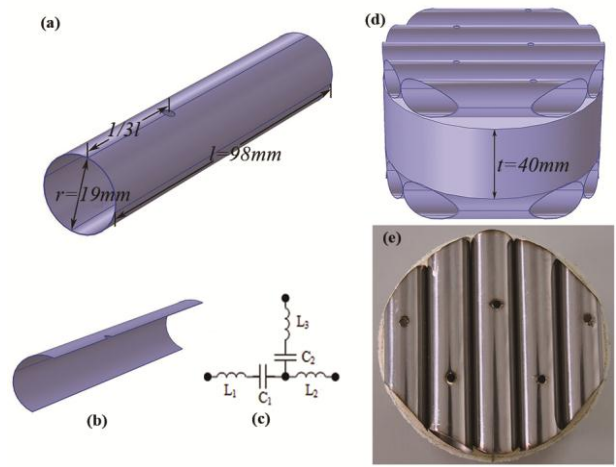

FIG. 1. Schematic diagrams of the PHST AM. (a) Three-dimensional schematic diagram, (b) cross section, and (c) equivalent LC circuit of the PHST. (d) Three-dimensional schematic and (e) physical views of the PHST AM.

Measurements of the transmission amplitudes and phase curves were conducted in a Shengwang transmission loss impedance tube apparatus,[17] which consisted of two impedance tubes, a loudspeaker at one end of the tubes, a four-channel data collecting analyzer, a power amplifier, and four microphones distributed at fixed positions along the tubes. The test sample was placed in the middle of the two hermetic tubes. The transmission amplitude and phase were acquired using the transfer function method, which processes the data measured by the microphones. Using a sound-absorbing material instead of a back tube in the transmission test system, a two-microphone measurement system was established. The complex reflection was also obtained using the transfer function method.

In the experiment, we fabricated sample A and plain sponge sample B. The transmission and reflection property of the samples are illustrated in Figs. 2(a) and 2(b). As the frequency increased from 400 Hz to 1930 Hz, the transmission ratio of the sponge matrix decreased from 7.3 dB to 4.7 dB. This phenomenon illustrates that the sponge is a non-dispersive sound medium and suitable for use as a sound matrix. The transmission ratio of the PHST sample became integrally lower, given that a distinct transmission peak was observed in the vicinity of 1706 Hz and an inverse phase near this position was found. At the same frequency, a reflection dip was found in the reflection curve, showing the basic form necessary to produce left-hand resonance conditions. The phase curve shown in Fig. 2(a) demonstrates a transmission peak corresponding to the negative phase change, which is an important characteristic of a metamaterial. Thus, according to Ref. 18, we conclude that the transmission peak is a metamaterial transmission peak.

Given that the wavelength of sound waves at a frequency of 1700 Hz in air is 200 mm, the lattice constant of sample A can be considered to be 38 mm. If the periodicity is considerably smaller than the corresponding longitudinal wavelength in the matrix, the metamaterial can be considered a homogeneous medium. Using the method for retrieving effective parameters from a homogeneous AM, the complex reflection coefficient was also obtained experimentally in the two-microphone measurement system, as shown in Fig. 2(b), which can be used with the complex transmission coefficient shown in Fig. 2(a) to calculate the effective bulk modulus and mass density of the PHST AM. From Ref. 19, as the incident acoustic wave is normal to the AM in our experiment, equations (1) and (2) can be simplified as follows:

$$R = \frac{1}{2}i\left(\frac{1}{Z_{eff}} - Z_{eff}\right)\sin(nkL) \cdot t, \text{ and} \tag{1}$$

$$T = \frac{2}{\cos(nkL)\left[2 - i\left(Z_{eff} + \frac{1}{Z_{eff}}\right)\tan(nkL)\right]}, \tag{2}$$

After inverse processing, the acoustic refractive index $n$ and impedance $Z_{eff}$ can be expressed as:

$$n = \pm\frac{1}{kL}\cos^{-1}\left[\frac{1}{2T}\left(1 - R^2 + T^2\right)\right] + \frac{2\pi m}{kL}, \text{ and} \tag{3}$$

$$Z_{eff} = \pm\sqrt{\frac{(1+R)^2 - T^2}{(1-R)^2 - T^2}}, \tag{4}$$

where $R$ and $T$ are the reflection and transmission coefficients, respectively, $k$ is the acoustic wave vector, $d$ is the thickness of the material, and $m$ is the branch number of the $\cos^{-1}$ function.

Using Eqs. (3) and (4), the effective bulk modulus and mass density can be obtained as

$$E_{eff} = {Z_{eff}}/{n}\,(E_0), \text{ and} \tag{5}$$

$$\rho_{eff} = n Z_{eff} \rho_0. \tag{6}$$

The effective bulk modulus and mass density from the experimental data were negative within the frequency ranges of 1648 Hz to 1718 Hz and 1100 Hz to 1930 Hz [see Figs. 2(c) and 2(d)]. The PHST AM had double-negative effective parameters, and its refractive index was also negative within the frequency range of 1648 Hz to 1718 Hz, as shown in Fig. 2(e).

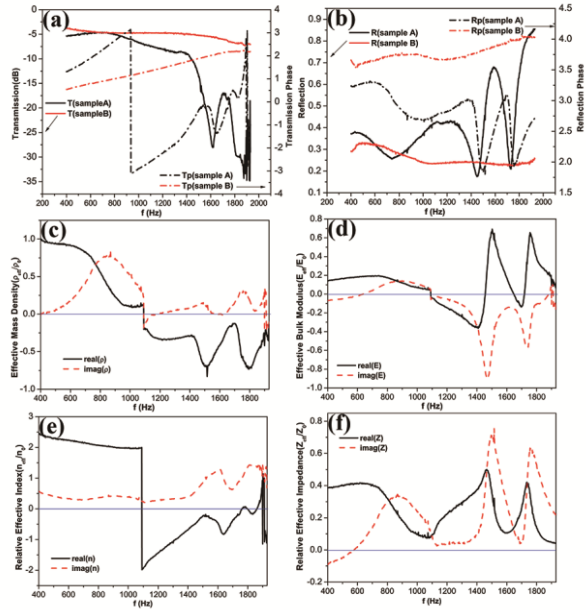

FIG. 2. (a) The transmitted amplitude and phase spectra of PHST AM. (b) The reflected amplitude and phase spectra of PHST AM. The (c) effective bulk modulus, (d) effective mass density of sample A. The real and imaginary parts of the (e) effective refractive index, (f) effective impedance of sample A

Negative refraction is the most significant singular characteristic of a metamaterial that distinguishes it from natural materials. A metamaterial has a flat lens function and can converge with an impinged divergent sound beam. An experiment was performed to check the possibility of point focusing by our PHST lens sample. We prepared a square PHST sample, as shown in Fig. 3(b). We constructed a 400 mm long, 400 mm wide, and 40 mm thick sponge matrix. The PHSTs were then arrayed on the upper and lower surfaces of the sponge matrix, as described above. We conducted a slab focusing experiment for the PHST square sample, as shown in Fig. 3(a). The speaker sound source was equivalent to a point source on one side of the sample. When the acoustic waves were transmitted through the sample, the sound field distribution along the *x*- and *y*-axes at the back of the sample was tested by a microphone and recorded by an oscilloscope. Fig. 3(a) presents the test coordinates, where the *x*-axis is perpendicular to the paper out, the origin of the coordinates for the central position of the square sample. We chose $f$=1.7 kHz in the double-negative frequency band and a wavelength of 200 mm. After testing in free space, we obtained a sound field distribution, as shown in Fig. 3(c). A sound focus spot 10 cm away from the sample was found, here, the sound intensity gradually increased and then gradually decreased about 10 cm behind the sample. Obviously, the distinct focus behavior was due to the PHST structure, while its focus spot intensity can attain to 10% and resolution can reach $\lambda/5$, breaking the diffraction limit. The result of the focus experiment further demonstrated the property of an AM based on a PHST sample.

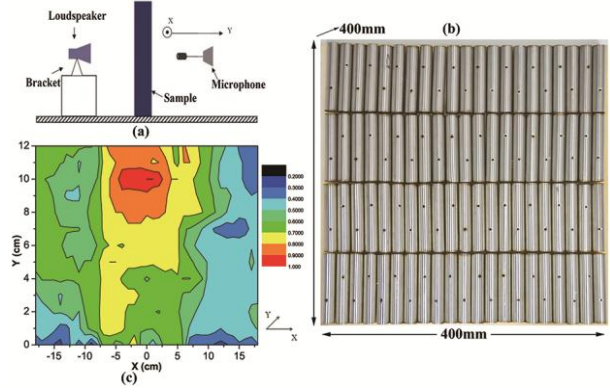

FIG. 3. (a) Schematic of the acoustic focus test. (b) Physical view of the square PHST sample. (c) Sound field distribution of the square PHST sample at 1.7 kHz in free space.

In conclusion, we proposed and experimentally studied a three-dimensional flute-model "molecular" structure with double-negative AM based on periodically arrayed PHSTs in a sponge matrix. Transmission experiments in the impendence tube system showed that the AM has a transmission peak and an inverse phase at 1706 Hz. Calculations of the effective bulk modulus and mass density of the PHST AM from the experimental data and resonant model demonstrated that the phenomena observed resulted from the local resonance of the PHSTs. The effective bulk modulus and mass density of the metamaterial were negative. Through PHST AM slab focusing experiments, the medium with a resonant structure was found to exhibit a distinct metamaterial property and its resolution reached $\lambda/5$, breaking the diffraction limit. This type of PHST AM can be used for negative refraction or acoustic cloaking.